\documentclass{article}

\usepackage{a4wide}
\usepackage{ipe}
\usepackage{amssymb}
\newcommand{\defaut}[1]{\underline{{#1}}}
\newcommand{\exces}[1]{\overline{{#1}}}
\newcommand{\region}[1]{{#1}}
\newcommand{\nvp}[2]{v_{\region{#1}}(#2)}
\newcommand{\vr}[2]{V_{\region{#1}}(#2)}
\newcommand{\bproof}{\noindent\textbf{Proof:~}}
\newcommand{\eproof}{\hfill $\Box$}

\newtheorem{theorem}{Theorem}
\newtheorem{lemma}[theorem]{Lemma}
\newtheorem{observation}[theorem]{Observation}
\newtheorem{corollary}[theorem]{Corollary}
\def\cgal{{\sc Cgal }}

\begin{document}



\title{
\vspace*{-15mm}
\fbox{\large published in Comp. Geometry Theory and Applications 33, 2005, p.3--17.}
\vspace*{5mm}\\
Inner and Outer Rounding of Boolean Operations on Lattice
Polygonal Regions}


\author{Olivier Devillers \and Philippe Guigue}
\date{
INRIA Sophia-Antipolis, BP93, 06902 Sophia Antipolis France. \tt http://www-sop.inria.fr/geometrica}

\maketitle

\begin{abstract}Robustness problems due to the substitution 
of the exact computation on real numbers by the rounded floating point
arithmetic are often an obstacle to obtain practical implementation 
of geometric algorithms.
If the adoption of the \emph{exact computation paradigm}~\cite{yd-ecp-95}
gives a satisfactory solution to this kind of problems
for purely combinatorial algorithms,
this solution does not allow to solve
in practice the case of algorithms that 
cascade the construction of new geometric objects.
In this report, we consider the problem of rounding 
the intersection of two polygonal regions
onto the integer lattice with inclusion properties.
Namely, given two polygonal regions $\region{A}$ and $\region{B}$
having their vertices on the integer lattice, 
 the inner and outer rounding modes
construct two polygonal regions $\region{A}\ \underline{\cap}\
\region{B}$ and $\region{A}\ \overline{\cap}\
\region{B}$ with integer vertices such that 
 $\region{A}\ \underline{\cap}\
\region{B} \subseteq \region{A}\ \cap\ \region{B} \subseteq
\region{A}\ \overline{\cap}\ \region{B}$.
We also prove interesting results on the Hausdorff distance, 
the size and the convexity of these polygonal regions.

{\bf Keywords:}
high level geometric rounding,
finite precision geometry,
intersection,
polygons
\end{abstract}


\section{Introduction}
Many geometric algorithms are designed in the Real RAM model,
and the use of rounded floating point arithmetic is well known
to create robustness problems: 
Numerical rounding errors done during the evaluation
of geometric predicates 
lead to inconsistent results and cause trouble in computer data
structures. 
The now classical solution of the  
{\em exact computation paradigm}~\cite{yd-ecp-95}
offers  an attractive solution 
for algorithms that do not construct new geometric objects
such as convex hulls or triangulations 
i.e whose results are purely
combinatorial 
(the position
of the points is not the result but the input of the algorithm).
The exact computation paradigm approach takes decisions on an exact
basis.
To achieve reasonably  efficient computation times  
this requires the use of well defined exact representations
of geometric objects: Typically, 
the coordinates of a point are assumed to be 
fixed size integers.

However, the exact computation paradigm is less satisfactory 
for algorithms that compute the geometric embedding of new objects. 
An intersection point between two line segments is a relevant
example of a construction of a new geometric object.
Such a point has rational coordinates and therefore is generally
not representable on the integer lattice.
If this point is used by the algorithm to make a decision,
 we must have an
exact representation of that point e.g. using rational numbers
or  implicit representation~\cite{fm-llook-00}
in order to ensure the exactness of that decision.
One drawback of this approach is that a constructed point does not 
use the original point representation
and thus in such a framework, algorithms cannot
be easily \emph{cascaded}, i.e. 
the (rational) output from one algorithm cannot be used as input for another
algorithm designed for usual input.\footnote{
In the sequel we will assume integer input for an algorithm using the
{\em exact computing paradigm} but it could also
be floating point or fixed point number (with fixed size representation)}

An alternative consists in rounding the constructions
that is 
replacing a geometric structure with arbitrary bit-length
coordinates by an approximating structure with (short) fixed 
bit-length coordinates.
However, rounding the coordinates of geometric objects like vertices of a
polygonal region is not straightforward 
since incidence information
may be invalidated by small perturbations of edges and vertices.
For instance, a polygonal region may be initially
convex or simple and can loose these properties after
a simple rounding of its vertices' coordinates.
Since these properties might be reused by other algorithms, this loss
of information is problematic.

Yet, there exist few published work in this direction, 
except for rounding line segment arrangements in the plane while preserving 
the topology of the 
arrangement~\cite{gy-frcg-86,gght-srlse-97,gm-rad-98,h-psifp-99,m-spgr-00}
(see Section~\ref{related_work}) 
and for rounding polyhedral subdivisions in three dimensions \cite{f-vrtdp-99}.
In this report, we are concerned with rounding the result of the
intersection of two planar lattice polygonal regions 
(i.e whose vertices have integer coordinates).
The result will extend trivially
to any other set operations on pairs of lattice polygonal regions.
Unlike the arrangement problem, we are 
interested in inclusion properties between the exact object 
and its rounded versions.
Previous works on arrangements can therefore not be used directly.
We propose in this report an algorithm which preserves
 such properties (see Figure~\ref{all-modes}).

Section 4 introduces the concept of inner and 
outer rounding of the intersection of lattice
polygonal regions. Section 5 deals 
with the practical  computation of these 
approximations. Section 6 proves
that a point on the boundary of a rounded version is at distance
less than $\sqrt{2}$ from the exact boundary,
and that convexity is somehow preserved.
Finally, Sections 7 and 8 generalize these rounding modes
to other set operations and to general polygonal regions.

\section{Related Work}
\label{related_work}

Three techniques for rounding line segments arrangements 
to a finite precision lattice have been proposed in the
literature.
All methods proceed by rounding the intersection points 
between the input line segments 
to their nearest lattice point.
Each original line segment is then replaced by 
a polygonal chain connecting the rounded version of
the endpoints and visiting all its rounded intersection points.
The techniques described below differ in the way that the polygonal chains
are constructed in order to guarantee metric and topological
properties\footnote{We refer the reader to \cite{gy-frcg-86} for an exhaustive inventory of shortcomings 
of the use of a simple rounding that maps each vertex 
of a line segments arrangement to its 
nearest
representable point.}.

\begin{figure}[!ht] \begin{center}
\IpeScale{48}
{\def\IPEfile{differentes_approches.ipe}\begingroup
  \catcode`\%=9\catcode`\!=0\catcode`\-=11\input{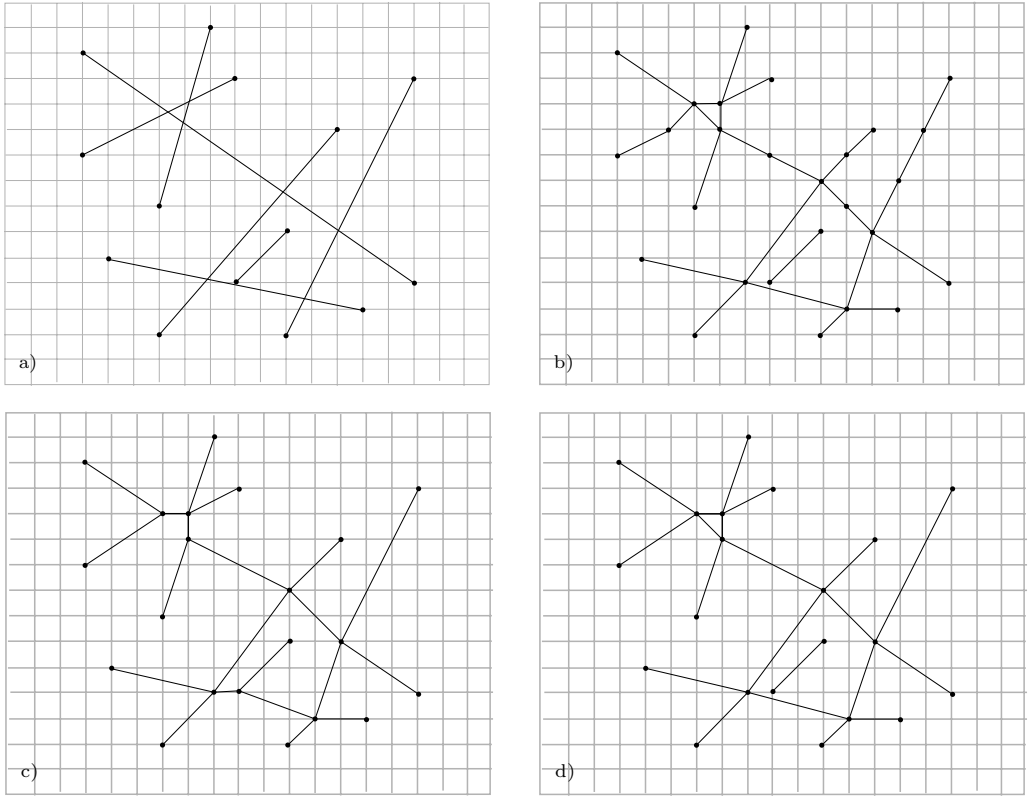}}
\caption{\label{RoundPrincipes}
a) A line segment arrangement, b) Its rounded version 
with the Greene-Yao perturbation technique, 
c) Its rounded version with the Snap Rounding technique, and
d) Its rounded version with the Shortest Path Rounding technique.}
\end{center}\end{figure}

\paragraph*{Greene-Yao perturbation technique}
The first method by Greene-Yao \cite{gy-frcg-86} treats each lattice
point as an obstacle and forbids 
any segment to go over an obstacle while its intersection points move to their
nearest lattice point.
Instead, an obstacle is introduced as a new vertex into 
the polygonal chain representing
the segment. The authors show that with this technique, 
edges move by a distance at most $\frac{\sqrt{2}}{2}$.
This algorithm has the disadvantage that it produces
  very fragmented polygonal
chains, which has an adverse effect on the efficiency of algorithms
and operations that use these fragmented line segments. 
Namely, this technique introduces $\Omega(\log|ab|)$ \emph{excess} lattice points 
onto a segment $ab$ where $|ab|$ denotes the length of the segment $ab$.

Later papers tried to reduce the number of additional vertices without 
introducing larger geometric derivations.

\paragraph*{Snap Rounding Paradigm}
Various researchers \cite{gght-srlse-97,gm-rad-98,h-psifp-99} have developed the Snap Rounding technique
for rounding line segments to the integer lattice.
The idea behind Snap Rounding is as follows. The plane is partitioned 
into pixels (i.e. isothetic unit squares)
centered at integer lattice points. 
A pixel is called \emph{hot} if it contains a vertex of the original
arrangement (that is either an endpoint or an intersection point of
the input segments).
The embedding is then rounded as follows:
Each original line segment is replaced by a polygonal chain 
that connects the centers of the hot pixels crossed by the segment.
This way, the number of vertices on
an edge is equal to the number of hot pixels crossed by the edge.

Guibas and Marimont \cite{gm-rad-98} give a very nice analysis of the 
properties of
Snap Rounding. 
One of its main properties is that it does not introduce 
any extra lattice points. Moreover, it can be easily shown that 
 the polygonal chain corresponding to an original segment is
contained within the Minkowski sum of the original segment with a
pixel (\emph{unit square}) centered at the origin.

\paragraph*{Shortest Path Rounding technique}
Shortest Path Rounding has been introduced by
Milenkovic \cite{m-spgr-00,m-pmsop-95} and introduces even fewer
additional incidences between the rounded segments
than Snap Rounding.
The basic idea is to round each intersection point to its nearest lattice point
and to replace each edge by the shortest path 
connecting the rounded endpoints 
that keeps all other rounded vertices at the correct side.
This technique has the advantage that it introduces minimum geometric
and combinatorial error (it gives the same result as the Snap
Rounding method in the worst case).
Moreover, unlike other finite precision geometric rounding techniques, 
Shortest Path Rounding can be applied to non uniform lattices.

Although these different techniques allow to preserve somehow
the topology of the exact arrangement, they
do not offer any inclusion or convexity guarantees
if they are applied on faces (and not only edges) of the arrangement.
The rounding modes proposed in this report 
are inspired from the presented methods
however they respond to the demand of such guarantees.

\section{Notations and preliminaries}

\begin{figure}[!ht]
\centering
\IpeScale{32}
{\def\IPEfile{modes_paper2.ipe}\begingroup
  \catcode`\%=9\catcode`\!=0\catcode`\-=11\input{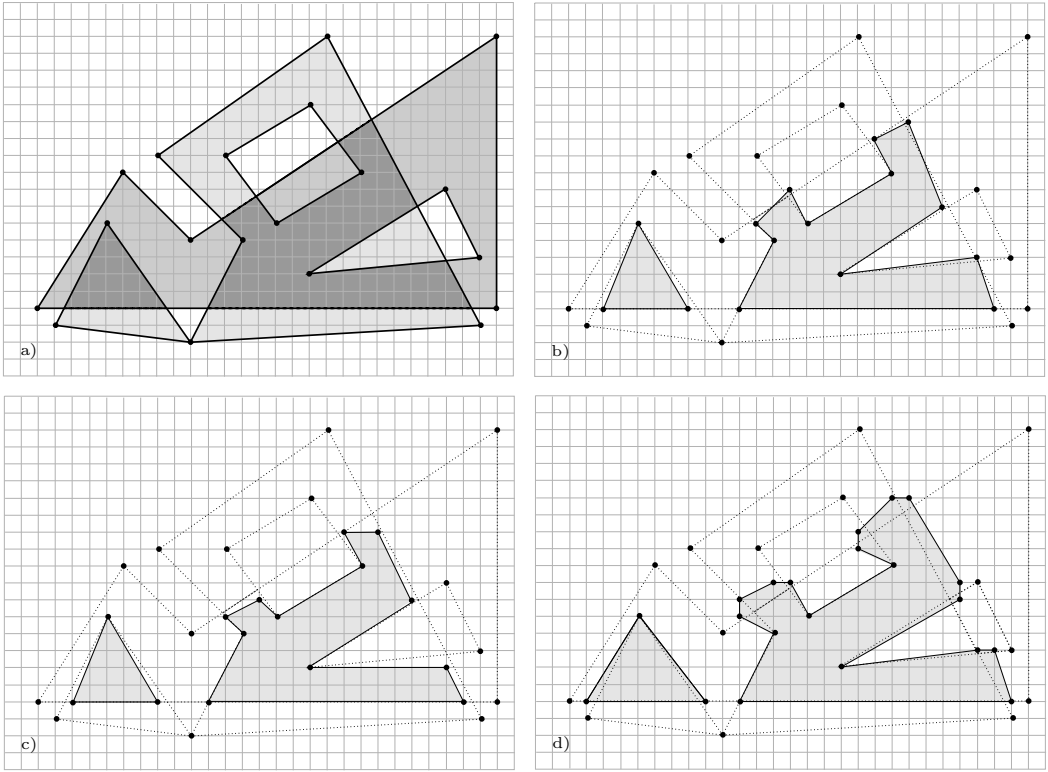}}
\caption{\label{all-modes}
a) The two input lattice polygonal regions
and their exact intersection region $\region{P}$.
b) The rounded version of $\region{P}$ with the Shortest Path 
Rounding technique. 
c) The inner rounding $\defaut{P}$. d) The outer rounding $\exces{P}$.}
\end{figure}

By a \emph{lattice point} or \emph{grid point}
we mean a point in $\mathbb{Z}^2$.  
A \emph{lattice polygon} is a polygon that defines 
a well defined interior and exterior (we allow a vertex to
coincide with another or to belong to an edge,
e.g. lowest vertex in Figure~\ref{all-modes}a)
 and whose vertices are lattice points. 
A \emph{lattice polygonal region} is a plane figure 
which can be expressed as a collection of lattice polygons 
having nested holes at any level of depth.
A lattice polygonal region has a well defined interior and exterior.
Our algorithms take such regions as input and give the output in the same form.

In the following, the complexity of a polygonal region $\region{P}$ 
defined as the number of
distinct vertices of $\region{P}$ is denoted by $|\region{P}|$.
The \emph{interior} of a polygonal region $\region{P}$, defined as the 
biggest open set contained in $\region{P}$, is denoted by $\region{P}^o$.
The \emph{boundary} of $\region{P}$ is denoted by $\partial{\region{P}}$.
We will say that a point $p$ belongs
to a polygonal region $\region{P}$, and note $p\in \region{P}$,  
if $p$ belongs either to the interior or to the boundary of $\region{P}$.
Finally, $\region{P}^C$ 
will denote the set complement of $\region{P}$. \\

Given two polygonal regions $\region{A}$ and $\region{B}$, the Hausdorff distance
$d_H(\region{A},\region{B})$ between $\region{A}$ and $\region{B}$ is defined as 
$$d_H(\region{A},\region{B})=max(d_h(\region{A},\region{B}),d_h(\region{B},\region{A}))$$ where $d_h(\region{A},\region{B})=\max_{a\in \region{A}}\min_{b\in
\region{B}}d(a,b)$ and $d(a,b)$ denotes the Euclidean distance
 between these points.\\

We will use the following definition of visibility.  
For two points $p$ and $q$ that belong to a polygonal region $\region{P}$,
we say that $q$ is \emph{visible} from $p$ within $\region{P}$,
if every point of the line segment $pq$ lies in $\region{P}$. 
The \emph{visibility region}, $V_P(p)$, of a point $p\in \region{P}$ 
is defined as the
locus of all points $q \in \region{P}$ that are visible from $p$.
The \emph{nearest visible lattice point} of $p$, denoted by $\nvp{P}{p}$,
is defined as the nearest grid point to $p$ that 
belongs to $\vr{P}{p}$ with any tie-breaking rule if $p$
is equidistant to several lattice points.
Finally, for a vertex $v \in \region{P}$ and an edge $e \in \region{P}$, 
we say that $v$ is 
\emph{vertically visible} from $e$, if it exists a vertical line segment
that connects $v$ to $e$ that is entirely contained in $\region{P}$.\\

We describe in the next section the scheme used to define the inner 
and the outer rounding of a polygonal region
and state the properties of the rounding scheme 
in the case where the input regions result 
from the intersection of two lattice polygonal regions.
Note that from the application of de Morgan's laws,
all set operations reduce to the 
complementary operation (whose computation is trivial)
and to the intersection operation.
Section \ref{set_operations} enumerates the properties 
satisfied when the exact region to be rounded comes from a union or a 
set difference operation.

\section{Rounding Modes}

\subsection{Inner Mode}

\label{inner_mode}
Suppose we start with two  input lattice
polygonal regions $A$ and $B$.
One can intuitively visualize the rounding process of 
the  polygonal region corresponding to the intersection 
of these two regions 
using the analogy used
by Greene and Yao \cite{gy-frcg-86}.
Look at the edges of this region as if they were rubber
bands rooted at their two endpoints and let every vertex of the
intersection
be marked by a rigid post.
These vertices may be vertices of $A$, $B$  or intersections between edges
of $A$ and $B$.
Each of these rigid post 
is then treated as an obstacle and we do not allow the rubber bands to
go over an obstacle .
Posts at original vertices of $A$ and $B$
are at lattice positions and remain fixed
while  posts at intersection of an edge of $A$ with an edge of $B$
move to their nearest visible lattice point inside $A\cap B$.
Now, if we release rigid posts that correspond to vertices that have
lost their convexity (vertices that were convex and became
concave), then the resulting polygonal region gives the inner rounded
polygonal region.

Theorem \ref{defaut_prop} states some properties of the obtained rounded
region in the case where $\region{P}$ corresponds to the intersection
of two lattice polygonal regions $\region{A}$ and $\region{B}$
(the proof is postponed until the Section \ref{propd}).

\begin{theorem}
\label{defaut_prop}
The inner rounding $\defaut{P}$ of $\region{P}=\region{A}\ \cap\ \region{B}$
satisfies the following properties:\\
1) $\defaut{P}$ is  lattice polygonal region,\\
2) $\defaut{P}$ is contained in $P$,\\
3) $d_H(\defaut{P}^C,(\region{P}^o)^C)<\sqrt{2}$,\\ 
4) $|\defaut{P}|\leq |\region{P}|$,\\
5) A concave vertex of $\defaut{P}$ does always correspond to a
concave vertex of $\region{P}$.

\end{theorem}

From property 5) we have the following corollary:
\begin{corollary}
If $\region{P}_i$ is a convex component of $\region{P}=\region{A}\
\cap\ \region{B}$ and if $\underline{\region{P}_i}$ is not empty
then $\underline{\region{P}_i}$ is a convex component of $\defaut{P}$.
\end{corollary}

\subsection{Outer Mode}

\label{outer_mode}
\begin{figure}[!ht]
\centering
\IpeScale{32}
{\def\IPEfile{modes3.ipe}\begingroup
  \catcode`\%=9\catcode`\!=0\catcode`\-=11\input{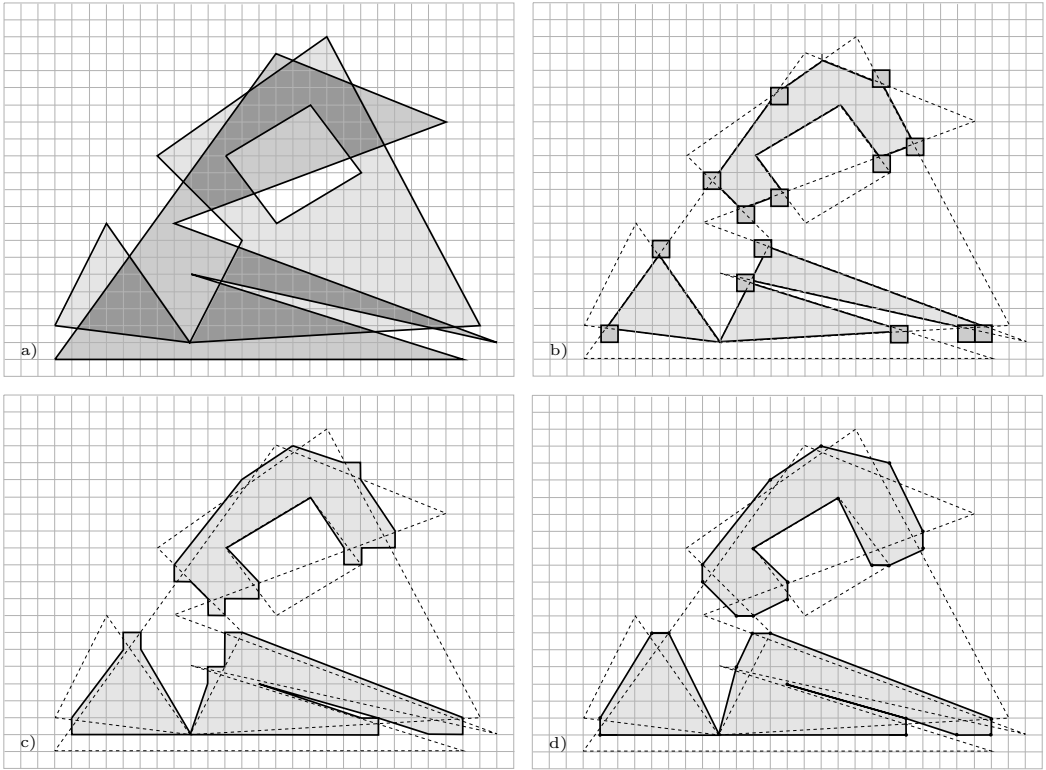}}
\caption{\label{outer_mode_algo}
a) The two input lattice polygonal regions.
b) Their exact intersection region $\region{P}$ and the set of pixels
$\region{I}$. 
c) The rounded version obtained by
computing ${({\region{P}}^C \underline{\cap}\ \region{I}^C)}^C$.
d) The region $\exces{P}$ obtained
by removing superfluous reflex vertices.}
\end{figure}

Given two  polygonal regions in the plane, 
the process leading to the computation of the outer rounding 
of their exact intersection region can be split in three steps.  
The idea is to bring the problem back to an inner
intersection computation (cf. Figure \ref{outer_mode_algo}).
To do so, the exact intersection region $\region{P}$ is first computed.
Then, for each vertex $v=(v_x,v_y)$ of $\region{P}$ that is not representable 
on the integer lattice is associated a pixel (unit square of the grid)
having respectively $(\lfloor v_x \rfloor, \lfloor v_y \rfloor)$ and 
$(\lceil v_x \rceil, \lceil v_y \rceil)$ as bottom left 
and top right vertex.\footnote{If 
one coordinate of $v$ is an integer but not the other,
then  the pixel degenerates into a unit segment, for simplicity the term pixel
in the sequel will include this kind of degenerate pixels.}
The outer rounding $\exces{P}$ of $\region{P}$ is then obtained
from $\region{P}$ and the set $\region{I}$ of all pixels containing 
non representable vertices of $\region{P}$ by carrying out
the operation $((\region{P}^C) \underline{\cap} (\region{I}^C))^C$.
A last pass removes all extraneous reflex vertices 
of the obtained polygonal region (see Section \ref{outer_intersection_algorithm}).

Theorem \ref{excess_prop} states some properties of the obtained rounded
region in the case where $\region{P}$ corresponds to the intersection
of two lattice polygonal regions $\region{A}$ and $\region{B}$
(the proof is postponed until the Section \ref{prope}).

\begin{theorem}
\label{excess_prop}
The outer rounding $\exces{P}$ of $\region{P}=\region{A}\ \cap\ \region{B}$
satisfies the following properties:\\
1) $\exces{P}$ is a  lattice polygonal region,\\
2) $\exces{P}$ contains $P$,\\
3) $d_H(\exces{P},\region{P})<\sqrt{2}$,\\ 
4) $|\exces{P}|< |\region{P}|+3k+h$, where $k$ is the number of
non-lattice vertices of $\region{P}$ and $h$ is the total number of
intersecting pairs between the edges of $\region{P}$ and those of $\region{I}$.\\
\end{theorem}

\section{Practical Algorithms} 

\label{algorithms}

From the analogy used in the Section \ref{inner_mode},
it is easy to see that each rigid post that corresponds to 
a vertex of  $\region{P}$ and that causes an edge 
of the intersection region to be broken during the movement of all posts 
corresponds to a reflex vertex of the exact intersection region.
Given two lattice polygonal regions $\region{A}$ and
$\region{B}$, the only vertices of the polygonal region 
 $\region{P}=\region{A}\ \cap\ \region{B}$ that are not representable onto
the integer lattice (that is the only vertices that need to be rounded)
correspond to the intersection points between an edge of $\region{A}$
and an edge of $\region{B}$.
From the definition of the intersection operation,
these non representable vertices can only form a convex vertex of
$\region{P}$.
Consequently, each reflex vertex of $\region{P}$ comes from
a reflex vertex of one of the two input regions and is therefore
a lattice vertex.

The algorithm for rounding the intersection of two lattice polygonal 
regions with the inner mode
is essentially based on the reflex vertical 
decomposition of the exact intersection region.
The purpose of the construction of this map is twofold:
1) It gives a convex decomposition of the original region that will
permit to avoid complex visibility calculation, 
2) It determines for each edge of the region 
a subset of the original vertices that 
should be snapped in order to avoid the introduction
of extraneous intersections. 

\subsection{The Reflex Vertical Decomposition}

\begin{figure}[!ht]
\centering
\IpeScale{45}
{\def\IPEfile{bool_vert.ipe}\begingroup
  \catcode`\%=9\catcode`\!=0\catcode`\-=11\input{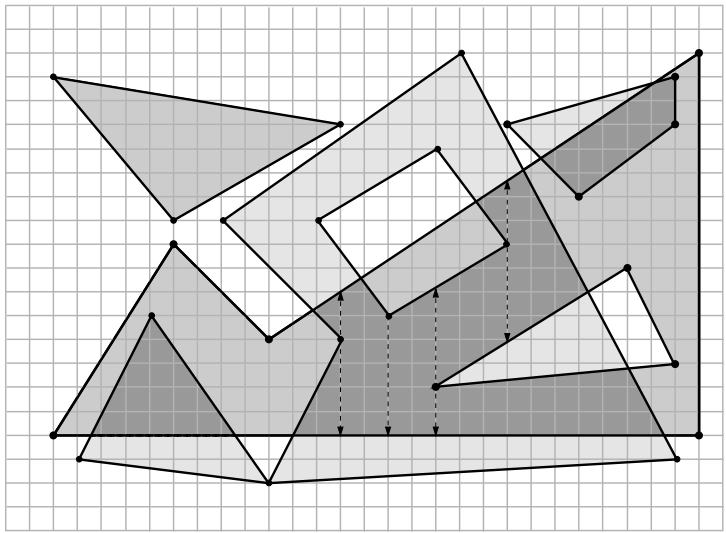}}
\caption{\label{fig-rvd}The reflex vertical decomposition of the 
intersection of two lattice polygonal regions.}
\end{figure}

The reflex vertical decomposition of a planar polygonal region 
is constructed by extending 
from each reflex vertex of the input region 
two vertical rays in the interior of the region 
in both the upward and downward directions.
These rays are the maximal vertical segments 
such that their relative 
interior does not intersect any edge of the polygonal region.
The reflex vertical decomposition of a polygonal region 
i.e. the subdivision of this region  induced by the edges of the
region and by the rays issued from its reflex vertices
is a partition of the input region into convex cells (see Figure \ref{fig-rvd}).

Before detailing the practical algorithm  we first prove
some properties of this decomposition.

\begin{figure}[!ht]
\centering
\IpeScale{110}
{\def\IPEfile{cellule_.ipe}\begingroup
  \catcode`\%=9\catcode`\!=0\catcode`\-=11\input{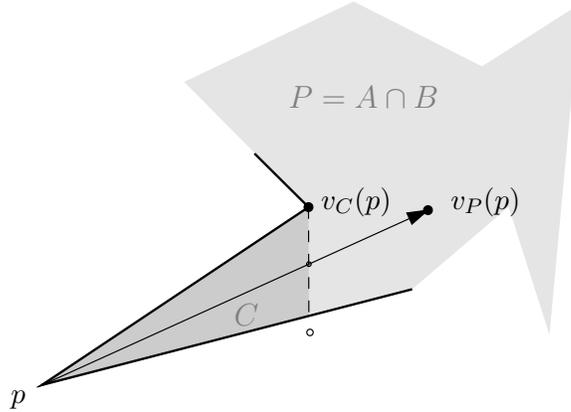}}
\caption{\label{cellule}If $p$ is a vertex of a convex cell
$\region{C}$ then $\nvp{P}{p} = \nvp{C}{p}$.}
\end{figure}

\begin{lemma}
\label{cell_lemma}
Given $\region{P}$ the exact intersection of planar lattice polygonal regions,
$p$ a vertex of $\region{P}$ and $\region{C}$ a convex cell of the 
reflex vertical decomposition of $\region{P}$ having $p$ as vertex then 
$\nvp{P}{p}=\nvp{C}{p}$.
\end{lemma}
\bproof
We prove this by contradiction. Suppose that 
$\nvp{P}{p}\neq\nvp{C}{p}$.
As $\nvp{P}{p}$ and $\nvp{C}{p}$ must be distinct
points $\nvp{P}{p}$ cannot belong to $\region{C}$. 
Therefore, the line segment connecting $p$ to $\nvp{P}{p}$ must cross 
the boundary of $\region{C}$ (cf. Figure \ref{cellule}). 
Since $\nvp{P}{p}$ is visible from $p$, the crossed 
boundary can only be a vertical wall emanating from a reflex vertex. 
Yet, this is impossible since, in this case, 
the two lattice points on the crossed ray immediately above 
and below the crossing are closer to $p$ than $\nvp{P}{p}$.
One of these two lattice points 
is between the crossing and the source of the ray and 
thus inside $C$ and visible by convexity of $\region{C}$.
This contradicts the fact that $\nvp{P}{p}$ cannot belong to $\region{C}$ and 
therefore the claim we made in the proof.
\eproof

\begin{figure}[!ht]
\centering
\IpeScale{95}
{\def\IPEfile{sslab.ipe}\begingroup
  \catcode`\%=9\catcode`\!=0\catcode`\-=11\input{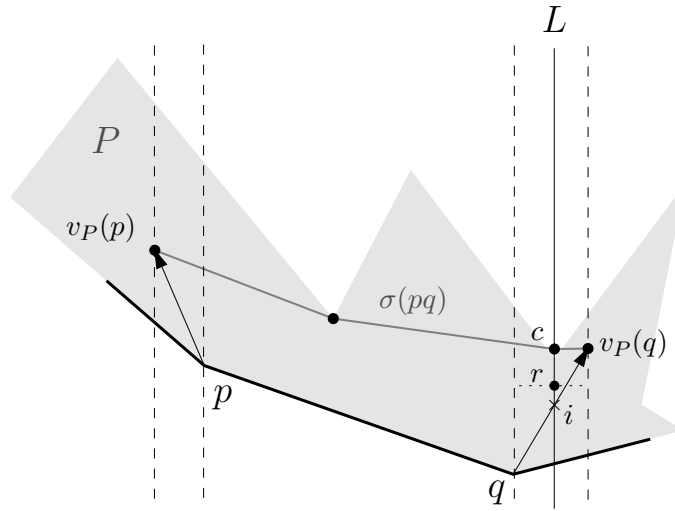}}
\caption{\label{slab}If $c$ is a vertex of $\sigma(pq)$ between 
$\nvp{P}{p}$ and $\nvp{P}{q}$ then $c$ is vertically visible 
from $pq$ in $\region{P}$.}
\end{figure}

\begin{lemma}
\label{slab_lemma}
Let $\region{P}$ be the exact intersection of planar lattice polygonal
regions, $pq$ be an  edge of $\region{P}$, 
$\defaut{P}$ be the inner rounding of $\region{P}$, and 
$\sigma(pq)$ be the polygonal chain connecting $\nvp{P}{p}$ to 
 $\nvp{P}{q}$ that corresponds to the rounded counterpart of
$pq$ in $\defaut{P}$.
The set of vertices of  $\sigma(pq)$ between $\nvp{P}{p}$ and
 $\nvp{P}{q}$ are reflex vertices of $\region{P}$ vertically visible
from $pq$ in $\region{P}$.
\end{lemma}
\bproof
By construction of the polygonal chain $\sigma(pq)$,
the vertices of $\sigma(pq)$  between $\nvp{P}{p}$ and
$\nvp{P}{q}$ necessarily correspond to reflex vertices of
$\region{P}$.
We show in the following that these vertices are  vertically visible
from $pq$ in $\region{P}$.
Here again, we prove this by contradiction.
Assume that there exists a vertex $c$ of  $\sigma(pq)$
between $\nvp{P}{p}$ and $\nvp{P}{q}$ such that $c$ is not vertically
visible from $pq$ in $\region{P}$.
Since $c$ belongs to $\sigma(pq)$ and is not vertically visible from 
$pq$ in $\region{P}$, $c$ surely lie in one of 
the two $x$-intervals induced by the segments $p\nvp{P}{p}$ and
$q\nvp{P}{q}$ (cf. Figure \ref{slab}).
Suppose wlog that $c$ belongs to the $x$-interval 
induced by the segment  $q\nvp{P}{q}$ and let $i$ be the point of
intersection between $q\nvp{P}{q}$ and the vertical line $L$ 
passing through $c$.
Since $c$ is a reflex vertex of $\region{P}$, $c$ must lie at a
lattice site and $L$ is a lattice vertical line.
But this is impossible, since in this case there exists a lattice
point $r$ on $L$ between $i$ and $c$ that is visible from $P$
and closer to $q$ than  $\nvp{P}{q}$, which contradicts 
the fact that $\nvp{P}{q}$ is the nearest visible lattice point from
$q$ in $\region{P}$.
\eproof

\subsection{Inner Intersection Algorithm}

\label{inner_algorithm}
Let $\region{A}$ and $\region{B}$ be two lattice polygonal regions in the plane.
The algorithm works in three steps.
The first step constructs the arrangement of the edges of $\region{A}$ and 
$\region{B}$ 
and computes the reflex vertical decomposition 
of the intersection region $\region{P}=\region{A}\ \cap\ \region{B}$ 
by the use of a Bentley-Ottmann-like
sweep line algorithm.

Based on this vertical decomposition, the second step 
rounds each vertex of $\region{P}$ 
that does not lie at lattice site to the nearest visible 
lattice point that belongs to its incident convex cell in the vertical
decomposition.
At the same time, each edge of $\region{P}$ is replaced by a polygonal chain
that connects its two rounded endpoints and passes through the set 
of all its vertically visible reflex vertices
in the order of their vertical projection on the edge.

The last step finally performs a variant of the  Graham's scan algorithm
for the  convex hull computation 
over the set of the resulting polygons (or holes).
This procedure removes all the reflex vertices 
from each polygon/hole except the ones corresponding to original reflex vertex 
(that is, it removes each reflex vertex that corresponds to a
 rounded intersection point or a visited vertically visible reflex vertex).

Given a vertex $v$ of  $\region{P}$ and its associated convex cell
$\region{C}$, the computation of the nearest  visible 
lattice point of $v$ in $\region{C}$ 
can be done  using the algorithm described in \cite{prisme-these-guigue}
in time $O(m\log m \log N)$ where
$m =  |\region{C}|$ and $N\times N$ is the size of the 
lattice containing $\region{C}$.
This algorithm, based on the continued fraction 
expansion technique, 
is inspired from the algorithm developed by H.S.Lee and R.C.Chang \cite{lc-avcpg-92}
which solves the problem in time $O(m + \log l)$,
where $l$ is the diameter of the convex cell.
However, this latter needs the use of an exact arithmetic 
on algebraic numbers to be implemented robustly 
(while our algorithm in \cite{prisme-these-guigue} can be 
implemented using exact evaluation of degree 4 polynomials
whose entries are integers: the coordinates of alttice points).

\begin{theorem}
The inner rounding $\defaut{P}$ of a region
$\region{P}=\region{A}\ \cap\ \region{B}$ can be computed in time $O((n+k)\log
n + k|\region{P}|\log |\region{P}|\log N)$ where $n$ is the total number of edges
of the two input regions, $k$ is the number of edges of $\region{A}$
and $\region{B}$ that intersect 
and $N\times N$ is the size of the lattice.

\end{theorem}
\bproof
Given the two input regions $\region{A}$ and  $\region{B}$, 
the reflex vertical decomposition of their intersection region
is a by-product of the trapezoidal map of their edges.
Therefore it can be calculated in time $O((n+k) \log n)$ 
where $n$ is the total number of edges of  $\region{A}$ and
$\region{B}$ and $k$ is the number of intersecting pairs. 
The second step of the algorithm computes at most $k$ nearest visible
lattice points in convex cells of size at most $|\region{P}|$ in time
$O(|\region{P}| \log |\region{P}| \log N)$ and produces, in the worst case,  
a set of polygons/holes having a total of $|\region{P}| + 2r$ vertices where
$r$ is the number of reflex vertices of $\region{P}$ (each reflex
vertices being vertically visible from at most two edges of $\region{P}$).
Given an edge of the intersection region and its two rounded
endpoints, its associated polygonal
chain can be constructed in time linear with the number of 
intersections between the edge and the vertical walls of the
decomposition
and thus can be done in time $O(|\region{P}|)$.
Putting all together and since $r<|\region{P}|$ and $k\leq |\region{P}|$ we obtain
a worst case complexity of $O((n+k) \log n + k|\region{P}|\log
|\region{P}|\log N)$ for the whole algorithm.
\eproof

\subsection{Outer Intersection Algorithm}

\label{outer_intersection_algorithm}
The algorithm for computing the outer rounding of the intersection
of two lattice polygonal regions is essentially based
on the algorithm of Section \ref{inner_algorithm}
and can be directly deduced from the description given in Section 
\ref{outer_mode}.
However, we discuss here a way to reduce 
the number of extraneous reflex vertices of $\exces{P}$,
namely the extraneous 
reflex vertices of  $\exces{P}$ issued from the vertices of the pixels 
of $\region{I}$, 
that derive from the straightforward computation  
of $\exces{P}$ as ${({\region{P}}^C \underline{\cap}\
\region{I}^C)}^C$.

Contrary to the inner rounding of an intersection region,
the outer rounding mode (as described in Section \ref{outer_mode}) 
does not offer any guaranty on the convexity/concavity
preservation  of the exact region's  vertices.
Some reflex vertices of $\region{P}$ can disappear in $\exces{P}$,
in the same manner some extraneous reflex vertices (that correspond to
vertices of $\region{I}$ and thus do not appear in $\region{P}$)
can appear in $\exces{P}$. 
A simple improvement consists in removing all reflex vertices appearing 
in $\exces{P}$ if they have no counterpart in $\region{P}$ and if
their removing does not produce any topological change.
Some precautions  must be taken in order to preserve a 
maximal distance between the points of $\exces{P}$ and the points 
of $\region{P}$ less than $\sqrt{2}$. 
A solution may consist in removing a reflex vertex $r$ of 
$\exces{P}$ only if there exists an edge $e$ of $\region{P}$
such that $r$, the vertex preceding and the vertex following
$r$ on $\exces{P}$'s boundary
all lie at a distance
less than $\sqrt{2}$ from $e$.
This kind of simplification permits a reduction of extraneous reflex
vertices of $\exces{P}$ of a factor $O(k)$ in the best case.
Moreover, this additional pass is sufficient to guarantee 
as a side-effect the following property:
If no components of $\region{P}$ are merged in $\exces{P}$
(that is if $\region{P}$ and $\exces{P}$ have exactly the same number
of polygons) then a convex component of  $\region{P}$ remains 
convex in $\exces{P}$.

\begin{theorem}
The outer rounding $\exces{P}$ of a region
$\region{P}=\region{A}\ \cap\ \region{B}$ can be computed in time $O((n+h)\log
n + kp\log p\log N)$ where $n$ denotes the total number of vertices
of the two input regions, $h$ denotes the number of
intersection points between the edges of $\region{A}$ and the edges of
$\region{B}$, 
$N \times N$ is the size of the lattice,
$k$ denotes the number of 
intersection points between an edge of  $\region{P}$ and an edge of 
the set of pixels $\region{I}$ and $p = \max(|\region{P}|,|\region{P}^C \cap \region{I}^C|)$.
\end{theorem}
\bproof
The computation of the exact intersection region $\region{P}$ 
can be done in time $O((n+h) \log n)$ where $h$ denotes the number of
intersection points between the edges of $\region{A}$ and the edges of
$\region{B}$. The computation of the reflex vertical decomposition
of $(\region{P}^C \cap \region{I}^C)$ can then be computed  in time 
$O((|\region{P}|+h+k)\log(|\region{P}|+h))$ where $k$ is the  
 number of intersection points between the edges of $\region{P}$ and the edges of
$\region{I}$.
Finally, the algorithm computes at most $k$ nearest visible  
lattice points in convex cells of size at most equals to
$m=|\region{P}^C \cap \region{I}^C|$ in time $O(m\log m \log N)$
using the algorithm described in \cite{prisme-these-guigue} and produces a set of
polygons/holes having a total number of vertices in $O(m)$.
The final step of the algorithm is linear in the number of vertices
of each polygons. Since $h \leq |\region{P}|$ and with $p =
\max(|\region{P}|,|\region{P}^C \cap \region{I}^C|)$, we obtain 
a worst case complexity of $O((n+h) \log n + kp\log p\log N)$ for the whole algorithm.
\eproof

\section{Proofs of Properties}

\subsection{Inner Intersection} 
\label{propd}

\begin{figure}[!ht]
\centering
\IpeScale{40}
{\def\IPEfile{inversion3.ipe}\begingroup
  \catcode`\%=9\catcode`\!=0\catcode`\-=11\input{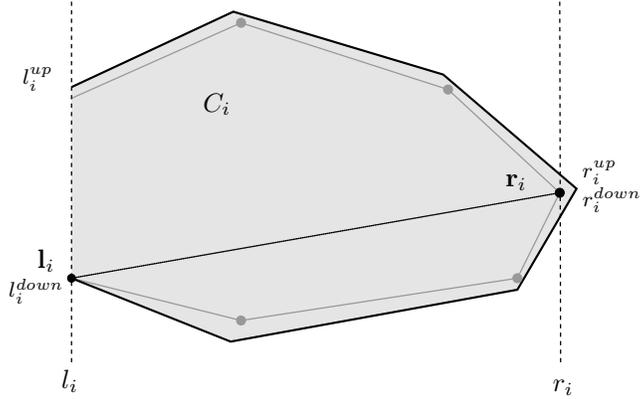}}
\caption{\label{inversion}The rounded counterparts of the polygonal
chains connecting $l_i^{down}$ to $r_i^{down}$ and 
$l_i^{up}$ to $r_i^{up}$ have opposite convexity and cannot invert
in $\region{C}_i$.}
\end{figure}

We now prove that the algorithm of Section \ref{inner_algorithm} 
computes an inner approximation of $\region{A}\ \cap \ \region{B}$ 
that satisfies the properties stated in
Theorem \ref{defaut_prop}. We first need the following lemmas:

\begin{lemma}
The computed polygonal region is a lattice polygonal region.
\label{valid}
\end{lemma}
\bproof
We prove that no extraneous intersections are introduced in 
the final approximation (though new incidences are permitted).
Let $\region{C}_{i=0..p}$ be the set of all convex cells of the 
vertical decomposition of the exact intersection. 
For each $\region{C}_i$, let $l_i$ and $r_i$ be the two vertical lines 
that pass through respectively the leftmost and the rightmost 
lattice point of $C_i$ (cf. Figure \ref{inversion}).
Now, if the intersection of $\region{C}_i$ with $l_i$ (resp. $r_i$) 
is a wall of $\region{C}_i$, let $l_i^{down}$ and $l_i^{up}$ 
(resp. $r_i^{down}$ and $r_i^{up}$) be the lower and the upper  
intersection of $l_i$ (resp. $r_i$) with $\region{C}_i$ and let
$\mathbf{l}_i$ (resp. $\mathbf{r}_i$) be the point on $l_i$ 
(resp. $r_i$) that corresponds to 
the reflex vertex from where the wall is stemming from.
Otherwise, let $l_i^{down}=l_i^{up}$ 
(resp. $r_i^{down}=r_i^{up}$)  
equal the leftmost (resp. the rightmost) 
vertex of $\region{C}_i$ 
and let $\mathbf{l}_i = \nvp{P}{l_i^{down}}$ (resp. $\mathbf{r}_i =
\nvp{P}{r_i^{down}}$).

The rounded counterparts of the polygonal chains 
connecting $l_i^{down}$ to  $r_i^{down}$, 
respectively $r_i^{up}$ to $l_i^{up}$), are convex (by convexity of the
original chains), therefore they are guaranteed not to lie 
above, respectively below, the edge  $\mathbf{l}_i \mathbf{r}_i$
and thus cannot invert in $\region{C}_i$.
\eproof

\begin{lemma}
All vertices of the computed polygonal region lie at lattice point
within the exact intersection region.
\label{lattice}
\end{lemma}
\bproof
There actually exist three types of vertices in the final
approximation: rounded intersection points, original input vertices
and snapped vertices corresponding to vertically visible input reflex
vertices. Since each intersection point rounds to its nearest visible
lattice point, the first type of vertex is guaranteed to lie at
lattice point within the intersection region. The two other types
of vertices correspond to lattice vertices of the exact intersection
region.
\eproof

\begin{observation}
\label{obs}
Given $pq$ an edge of the exact intersection region, the polygonal
chain $\sigma(pq)$ that connects $\nvp{P}{p}$ to $\nvp{P}{q}$
and corresponds to the rounded counterpart of $pq$ in $\defaut{P}$
is entirely contained in $\region{P}$ by construction.
\end{observation}

\begin{figure}[ht!]
\begin{center}
\IpeScale{63}{\def\IPEfile{proof1.ipe}\begingroup
  \catcode`\%=9\catcode`\!=0\catcode`\-=11\input{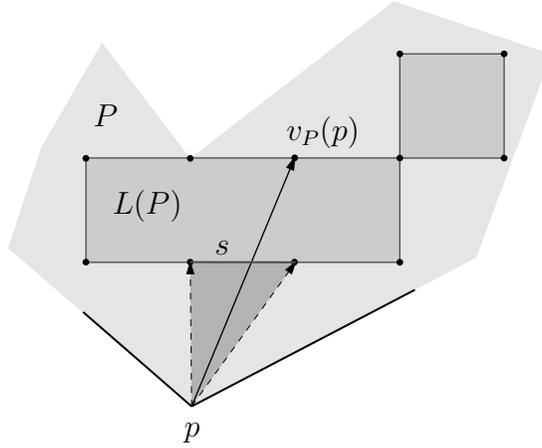}}
\caption{\label{f1}The segment $p\nvp{P}{p}$ cannot intersect the interior of $L$.}
\end{center}
\end{figure}

\begin{lemma}
\label{lemmaL}
Let $p$ be a vertex of $P$ and 
$L(P)$ be the union of all lattice points, unit lattice segments
and unit lattice squares that belong to the interior or to the 
boundary of $P$.
The segment connecting $p$ to $v_{P}(p)$ cannot intersect
the interior of $L(P)$.
\end{lemma}
\bproof
To intersect the interior of $L(P)$, the segment $pv_{P}(p)$ 
must intersect the interior of a unit lattice segment $s$ of
$\partial{L(P)}$ (cf. Figure \ref{f1}).
The two endpoints of $s$ are necessarily closer to $p$ 
than $v_{P}(p)$ and therefore cannot be visible from $p$
since they 
correspond, by definition of $L(P)$,
to lattice points that lie inside $P$.
Consequently,  
the relative interior of the segments connecting $p$ to these endpoints
must intersect the boundary of $P$.
But this is impossible since by definition 
both segments $p v_{P}(p)$ and $s$ cannot
intersect in their interior the boundary of $P$
and there cannot exist any visible reflex
(lattice) vertex of $P$ inside the triangle having $p$ and 
the two endpoints of $s$ as vertices since all points 
of this triangle are closer to $p$ than $v_{P}(p)$.
\eproof

\begin{lemma}
Given two lattice polygonal regions $\region{A}$ 
and $\region{B}$ of the plane, $d_H((\region{A}\ \underline{\cap}\ 
 \region{B}\ )^C , (\region{A}^o \cap \region{B}^o)^C)<\sqrt{2}$.
\label{hausdorff}
\end{lemma}
\bproof
From Observation \ref{obs}, $\defaut{P}=\region{A}\underline{\cap}\region{B}$
is included in $\region{P}=\region{A}\cap \region{B}$ and the
directional Hausdorff distance $d_h({(P^o)}^C,\underline{P}^C)$ is zero.
We prove in the following that all points of $P \setminus \underline{P}$
are at a distance less than $\sqrt{2}$ from $\partial{P}$. 
Let $pq$ be an edge of $P$ and let $E_{pq}$ be the polygon obtained by
appending the rounded chain $\sigma(pq)$ to $qp$. Notice that
$(\bigcup_{pq\in P} E_{pq})$ partitions $P \setminus \underline{P}$
except for the polygons $p_i$ of $P$ that do not have any rounded
counterpart. The bound is trivially proven for the latter polygons 
since they do not contain any lattice point. 
For the non-trivial case, we conclude that it is sufficient to prove 
that $d_h(E_{pq},\partial{P})<\sqrt{2}$
for any edge $pq$ of $P$.

Let $L(P)$ be the union of all lattice points, unit lattice segments
and unit lattice squares that belong to the interior or to the 
boundary of $P$.
By definition of $L(P)$, all points of $P \setminus {L(P)}^o$ are at a distance
less than $\sqrt{2}$ of $\partial{P}$, we therefore suppose in the
following that $E_{pq}$ is not entirely included in $P \setminus
{L(P)}^o$. 

\begin{figure}[ht!]
\begin{center}
\IpeScale{65}{\def\IPEfile{proof3.ipe}\begingroup
  \catcode`\%=9\catcode`\!=0\catcode`\-=11\input{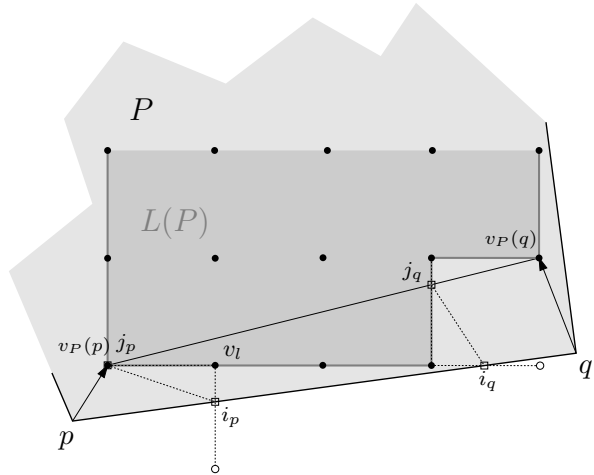}}
\caption{\label{f2}The polygons $pi_pj_pv_{P}(p)$ and $j_qi_qqv_{P}(q)$
are contained in $P \setminus L^o$ and the polygon $i_pi_qj_qj_p$ is
at a distance less than $\sqrt{2}$ to $pq$.}
\end{center}
\end{figure}

By lemma \ref{lemmaL}, the segments $pv_{P}(p)$ and $qv_{P}(q)$ 
cannot intersect the interior of $L(P)$ thus for $E_{pq}$ to intersect 
${L(P)}^o$, $\sigma(pq)$ must necessarily intersect ${L(P)}^o$.
Moreover, by convexity of the chain $\sigma(pq)$, there must
exist in this case at least one (lattice) vertex $v$ 
different from
$v_{P}(p)$ and $v_{P}(q)$ that lies in or on the boundary of $E_{pq}$.
Suppose wlog that $pq$ is oriented from left to right
with a positive or zero slope 
and that the interior of $P$ lies above $pq$. Finally, let $v_l$ the
$xy$-smallest point (w.r.t the lexicographic order) of the set $S$ of all
lattice points 
different from $v_{P}(p)$ and $v_{P}(q)$ 
that lie in or on the boundary of $E_{pq}$ (cf. Figure \ref{f2}).

Using the same arguments as in the proof of lemma \ref{slab_lemma}
and since $v_l$ is the $xy$-smallest point of $S$,
it is easy to show that $v_l$ is vertically visible from $pq$
and that  
the  vertical unit lattice segment having $v_l$ as top
vertex surely intersects $pq$ in a point $i_p$.
Similarly, since $pq$ is oriented from left to right and has a
positive or zero slope 
and since by lemma \ref{lemmaL}
the segments $pv_{P}(p)$ and $qv_{P}(q)$ 
cannot intersect the interior of $L(P)$, the horizontal unit lattice
segment having $v_l$ as right vertex surely intersects
$\sigma(pq)$ in a point $j_{p}$. Notice that, by construction, 
both $i_p$ and $j_p$ belong to the boundary of a same unit 
lattice square so that $\|i_pj_p\|<\sqrt{2}$.

Replacing $p$ by $q$ and 
applying a symmetry operation
on $E_{pq}$ such that $qp$ is oriented from left to right 
and has a positive or zero slope
with the interior of $P$ above $qp$,
we define similarly
two points $i_q$ and $j_q$ on $pq$ and $\sigma(pq)$ 
such that $\|i_qj_q\|<\sqrt{2}$.
We conclude that $d_h(E_{pq},\partial{P})<\sqrt{2}$
since the polygons $pi_pj_pv_{P}(p)$ and $j_qi_qqv_{P}(q)$
are contained in $P \setminus {L(P)}^o$ (by definition of $v_l$) and 
the polygon $i_pi_qj_qj_p$ is contained 
in the Minkowski sum of $i_pi_q$ with the interior of a disc of radius
$\sqrt{2}$ centered at the origin  (cf. Figure \ref{f2}),
and by convexity of $\sigma(pq)$, the portion of $\sigma(pq)$ between
$j_p$ and $j_q$ is included in $i_pi_qj_qj_p$. 
\eproof

We are now able to prove Theorem \ref{defaut_prop}.
Proof of property 1) comes from the combination of Lemmas \ref{valid}
and \ref{lattice}.
From 1) and by construction of the approximation the proof 
of Property 2) is trivial. 
Property 3) is proven in Lemma \ref{hausdorff}.
Proof of property 4) comes from the fact that each intersection point
rounds to at most one lattice point and that all extra vertices that appear
in the approximation correspond to original reflex vertices of the
exact intersection. 
The number of vertices of the final approximation is larger than the
number of vertices of the original region only when 
a vertex of $\region{P}$ is used several times in the approximation.
Property 5) is a direct consequence of the last step of the algorithm
since the convex-hull pass guarantees that no extra reflex vertices are
introduced in the final approximation.

\subsection{Outer Intersection} 
\label{prope}
In this section, we introduce some lemmas 
needed for the proof of Theorem \ref{excess_prop}.
Notice that property 3) cannot be deduced from Lemma \ref{hausdorff} 
since we must bound 
the distance between the points of $\exces{P}$
to  the \emph{exact} intersection region $\region{P}$
and not only to the region $(\region{P}^C \cap
\region{I}^C)^C$.
That is, we must exclude 
that there exist points of $\exces{P}$
that are close to a pixel of $\region{I}$
but at a distance greater than $\sqrt{2}$ from the region $\region{P}$.

\begin{lemma}
Given two lattice polygonal regions $\region{A}$ 
and $\region{B}$ of the plane, $d_H((\region{A}\ \overline{\cap}\ 
 \region{B}\ ) , (\region{A} \cap \region{B}))<\sqrt{2}$.
\label{hausdorff2}
\end{lemma}
\bproof
Since $\region{P}=(\region{A} \cap \region{B})$ is included in 
$\exces{P}=(\region{A}\ \overline{\cap}\  \region{B}\ )$, 
the (directional) Hausdorff 
distance from $\region{P}$ to $\exces{P}$ is zero.
Therefore, it is sufficient to show that each point of
$\exces{P} \setminus \region{P}$ is at a distance less than 
$\sqrt{2}$ to $\region{P}$.

Note that since no pixel of $I$ contains a lattice point
in its interior, the union $L({P}^C \cap {I}^C)$ of all the lattice points, lattice
segments and pixels that belong to the interior or to the boundary of 
$(\region{P}^C \cap \region{I}^C)$ is also the
union of all the lattice points, lattice
segments and pixels that belong to the interior or to the boundary of
$\region{P}^C$.
Therefore, if the polygon $E_{pq}$ (as defined in the proof of Lemma
\ref{hausdorff}) is contained in $(\region{P}^C \cap \region{I}^C)
\setminus {L({P}^C \cap {I}^C)}^o$ it is also contained in 
$\region{P}^C \setminus {L(P^C)}^o$, and 
$E_{pq}$ surely lies at a distance less than $\sqrt{2}$ to the boundary of $\region{P}$.

\begin{figure}[!ht]
\begin{center}
\IpeScale{45}
{\def\IPEfile{pp.ipe}\begingroup
  \catcode`\%=9\catcode`\!=0\catcode`\-=11\input{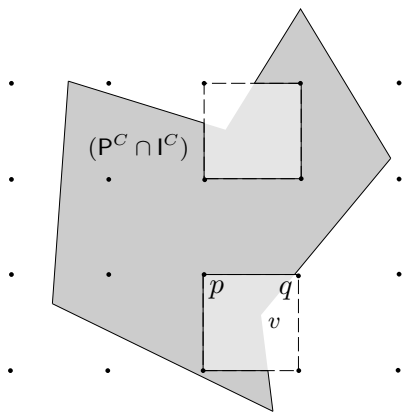}}
\caption{\label{p21}If $pq$ is issued from an edge of $I$ 
and at least one of its endpoints is a lattice point
then $\sigma(pq) \subseteq pq$ (in the example $\sigma(pq) = p$) and $\sigma(pq)$ lies at a distance
less than $\sqrt{2}$ to a vertex $v$ of $P$.}
\end{center}
\end{figure}

Otherwise, 
with the same arguments as in the proof of Lemma \ref{hausdorff},
we show that the part of $E_{pq}$ which is intersected by the interior of
${L(P^C)}^o$ surely lies at a distance less than $\sqrt{2}$ to $pq$.
Therefore if $pq$ is issued from an edge of $P$ then
the bound is trivially proven.
Otherwise, that is if $pq$ is issued from a pixel $Q$ of $I$, 
using the same arguments
as in the proof  of Lemma \ref{hausdorff}, there must exist a lattice
line passing through $v_l$ (as defined in the proof)
that intersects $pq$.
This line cannot intersect the relative interior of $pq$ since 
the edge $pq$ is included or equals a unit lattice segment.
Moreover, if  this line intersects $pq$ in one of its endpoints
then the intersected endpoint is necessarily a lattice point 
and therefore the rounded counterpart $\sigma(pq)$ of $pq$ is included
in $pq$ (cf. Figure \ref{p21}).
We conclude in this case that $d_h(E_{pq},\partial{P})<\sqrt{2}$
since there must exist a vertex of $P$ in the pixel $Q$
(namely, the vertex of $P$ that causes the presence of $Q$ in $I$).
\eproof

\begin{lemma}
\label{exces_bound}
The rounded region $\exces{P}$ of
$\region{P}=\region{A}\cap\region{B}$
has less than $|\region{P}| + 3k + h$ distinct
vertices where $k$ is the number of
non-lattice vertices of $\region{P}$ and $h$ is the total number of
intersecting pairs between the edges of $\region{P}$ and those of $\region{I}$.
\end{lemma}
\bproof
Since $\exces{P}$ corresponds to the complementary of the inner
rounding $\underline{\textsf{P}_{\region{I}}}$ of $\region{P}_{\region{I}} = (\region{P}^C \cap \region{I}^C)$, we have
from Theorem \ref{defaut_prop} that 
$|\overline{{\textsf{P}}_{\region{I}}}| \leq |\region{P}_{\region{I}}|$, and the number of vertices of
$\overline{\textsf{P}_{\region{I}}}$ is bounded by $|\region{P}_{\region{I}}|$.
If $\region{P}$ has a total of $n$ vertices and has $k$ vertices 
which are not representable on the integer lattice, $|\region{I}|\leq 4k$ and
$\region{P}_{\region{I}}$ has at most $(n-k)+4k$ lattice vertices and $h$ non
integer vertices where $h$ denotes the number of intersection point 
between $\region{P}$ and $\region{I}$ edges.
\eproof

Although the number of vertices of $\region{P}_{\region{I}}$ 
used as an upper bound on the complexity of $\exces{P}$ can be in the 
worst case in $O(nk)$, 
an additional pass of the algorithm can be used to
guarantee a total number of vertices of $\exces{P}$ 
which is linear in the number of vertices of the exact region
$\region{P}$.
More precisely, we show in \cite{prisme-these-guigue} that
the removal of all zero-area components (that is polygons or holes of 
$\exces{P}$ that have no interior)
 from the obtained region
allows to bound the worst case number of 
distinct vertices of $\exces{P}$ by $2n+3k$
without affecting the geometric error bound.
In addition, experimental results 
obtained with an implementation 
of the algorithm using the C++ library \cgal~\cite{prisme-cgal-3.0}
indicate that the number of 
additional vertices of $\exces{P}$ is very small in practice.

From the above lemmas, we are now able to prove Theorem \ref{excess_prop}.
The proof of property 1) and 2) can be directly deduced 
by construction of $\exces{P}$ from Theorem
\ref{defaut_prop}.
Property 3) is proved in Lemma \ref{hausdorff2}.
Property 4) is proved in Lemma \ref{exces_bound}.

\section{Rounding Set Operations}

\label{set_operations}

Theorems \ref{exces_prop_union} and \ref{defaut_prop_union}
enumerate the set of properties satisfied when the exact region $\region{U}$
comes from a union operation i.e. when $\region{U}=\region{A}\cup\region{B}$.
These properties can be directly obtained
from Theorems \ref{defaut_prop} and \ref{excess_prop} 
by replacing $\region{A}$ and $\region{B}$ by their 
complementary sets.

\begin{theorem}
\label{exces_prop_union}
The outer rounding $\exces{U}$ of $\region{U}=\region{A}\cap\region{B}$
satisfies the following properties:\\
1) $\exces{U}$ is a  lattice polygonal region,\\
2) $\exces{U}$ contains  $U$,\\
3) $d_H(\exces{U},\region{U})<\sqrt{2}$,\\ 
4) $|\exces{U}|\leq |\region{U}|$,\\
5) A convex vertex of $\exces{U}$ does always correspond to a
convex vertex of $\region{U}$.

\end{theorem}

\begin{theorem}
\label{defaut_prop_union}
The inner rounding  $\defaut{U}$ of $\region{U}=\region{A}\cup\region{B}$
satisfies  the following properties:\\
1) $\defaut{U}$ is a  lattice polygonal region,\\
2) $\defaut{U}$ is contained in $U$,\\
3) $d_H(\defaut{U}^C,(\region{U}^o)^C)<\sqrt{2}$,\\ 
4) $|\defaut{U}|\leq |\region{U}|+k+h$,where $k$ is the number of
non-lattice vertices of $\region{U}$ and $h$ is the total number of
intersecting pairs between the edges of $\region{U}$ and those of $\region{I}$.\\
\end{theorem}

The result for the set difference operation can equally be deduced
from Theorems 1, 2, 5 and 6
for each rounding mode.

\section{Rounding General Regions}

From the lemmas and algorithms presented so far in this report, the
inner/outer rounding of a general polygonal region
(for which we do not have any 
assumption on the representation of its vertices, for example
a region issued from a rotation operation)
can be obtained in the
following manner.
Let $\region{P}$ be a general polygonal region and consider 
$V_c$ (resp. $V_r$), the set of its convex (resp. reflex) vertices 
that do not lie at lattice sites.
Let now $\region{I}_c$ (resp. $\region{I}_r$) be the set of 
unit lattice squares that contain the vertices of $V_c$ (resp. $V_r$),
i.e. the set of quadrilaterals having respectively 
$(\lfloor v_x \rfloor, \lfloor v_y \rfloor)$ and 
$(\lceil v_x \rceil, \lceil v_y \rceil)$ as bottom left 
and top right vertex where $v=(v_x,v_y)$ is a vertex of $\region{I}_c$
(resp. $\region{I}_r$).

We define the inner rounding $\defaut{P}$ of $\region{P}$ as the
result of the rounding with the inner mode of the intersection of $\region{P}$ and
$\region{I}_r$, that is $\defaut{P}=\region{P}\ 
\underline{\cap}\  \region{I}_r$. 
Similarly, we define the outer
rounding $\exces{P}$ of $\region{P}$ as the complementary of the
rounding with the inner mode of the intersection of 
 $\region{P}^C$ and ${\region{I}_c}^C$, that is 
$\exces{P}=(\region{P}^C \underline{\cap}\ {\region{I}_c}^C)^C$.

The inner rounded intersection operations used in these definitions 
can be  directly  computed from the algorithm presented in Section 
\ref{algorithms}.
We remark indeed that all reflex vertices of the intersection regions 
$(\region{P}\ \underline{\cap}\ \region{I}_r)$ and 
$(\region{P}^C\ \underline{\cap}\ {\region{I}_c}^C)^C$
lie at lattice sites which is a sufficient condition to satisfy the properties 
of the reflex vertical decomposition stated in Lemmas 
\ref{cell_lemma} and \ref{slab_lemma}
and thus to prove the correctness of the algorithms.

We notice however that the absence of lattice segments supporting the
edges of the input region $\region{P}$ requires the use  
of a well suited number type and arithmetic in order 
to evaluate the numerical primitives that appear in the algorithm.
A number type and an arithmetic allowing the manipulation of algebraic 
numbers can be necessary for example if the input region $\region{P}$
are issued from a rotation operation.

The properties satisfied by the rounded regions $\defaut{P}$ and
$\exces{P}$ can be directly derived from the lemmas of the previous
sections. More precisely, the inner rounding $\defaut{P}$ is
a  lattice polygonal region contained in
$\region{P}$  such that $d_H(\defaut{P}^C,{\region{P}^o}^C)<\sqrt{2}$.
Moreover, if it exists, the rounded counterpart of a convex vertex of
$\region{P}$ is a convex vertex of  $\defaut{P}$. Finally, 
$\defaut{P}$ has less than $|\region{P}| + r + h_r$ distinct vertices 
where $r$ is the number of reflex vertices of $\region{P}$ and $h_r$
is the number of intersections between the edges of $\region{P}$ and the
edges of $\region{I}_r$.
The outer rounding $\exces{P}$ of $\region{P}$ is a 
 lattice polygonal region that contains
$\region{P}$ such that $d_H(\exces{P},\region{P})<\sqrt{2}$
and $|\exces{P}|<|\region{P}|+c+h_c$ where $c$ denotes the number of 
convex vertices of $\region{P}$ and $h_c$ denotes the 
 number of intersections between the edges of $\region{P}$ and the
edges of $\region{I}_c$.

We finally remark that the number of vertices needed to represent 
$\exces{P}$ can be reduced with the same kind of technique 
as described at the end of Section \ref{prope}.

\section{Conclusion}

We have given methods for computing the inner/outer 
rounding of the result of set operations 
on  two lattice polygonal regions in the plane.
The guarantees that the exact result of such operations 
contains (or is contained in) 
its finite precision approximation 
allows to introduce the geometric analogue of interval
arithmetic provided by the certified rounding modes
of the IEEE 754 norm for floating point arithmetic
operations.
The computation of such geometric intervals 
with respect to the inclusion relation
permits 
in particular to cascade various
geometric constructions as set operations, convex hulls or rotations
with a control on their bit complexity.
This result is a first step towards the definition of 
a complete system for performing rounded operations on polygonal
and polyhedral objects
 which would be of great
practical interest in many CAD applications.


\end{document}